\documentclass[fleqn,usenatbib]{mnras}

\usepackage{newtxtext,newtxmath}
\usepackage[T1]{fontenc}
\usepackage{ae,aecompl}

\usepackage{graphicx}	% Including figure files
\usepackage{amsmath}	% Advanced maths commands
\usepackage{amssymb}	% Extra maths symbols
\usepackage{newtxtext,newtxmath}
\usepackage[T1]{fontenc}
\usepackage{ae,aecompl}
\usepackage{bm}

\newcommand{\Msun}{\>{\rm M_{\odot}}}
\newcommand\degrees{^\circ}

\def\aj{AJ}% Astronomical Journal
\def\apj{ApJ}% Astrophysical Journal
\def\apjl{ApJ}% Astrophysical Journal, Letters
\def\apjs{ApJS}% Astrophysical Journal, Supplement
\def\aap{A\&A}% Astronomy, and Astrophysics
\def\mnras{MNRAS}% Monthly Notices of the RAS
\def\pasp{PASP} % Publications of Astronomical Society of the Pacific
\def\nat{Nature}% Nature
\title[Drivers of disc tilting I]{Drivers of disc tilting I: Correlations and possible drivers for Milky Way analogues}

\author[Earp et al.]{Samuel W. F. Earp$^{1,2}$\thanks{E-mail: swfearp@gmail.com},
Victor P. Debattista$^{2}$,
Andrea V. Macci\`{o}$^{3,4}$,
\newauthor Liang Wang$^{5}$,  
Tobias Buck$^{4}$ and Tigran Khachaturyants$^{2}$\\
$^1$ Sertis Corporation, 597/5 Sukhumvit Road, Watthana, Bangkok, 10110, Thailand \\
$^2$ Jeremiah Horrocks Institute, University of Central Lancashire, Preston, PR1 2HE, UK \\
$^3$ New York University Abu Dhabi, PO Box 129188, Saadiyat Island, Abu Dhabi, United Arab Emirates \\
$^4$ Max-Planck-Insitute for Astronomy, K\"{o}nigstuhl 17, D-69117 Heidelberg, Germany \\
$^5$ University of Western Australia, Crawley, WA 6009, Australia\\
}

\date{Accepted 2019 July 23. Received 2019 July 22; in original form 2018 November 16}

\pubyear{2019}

\begin{document}
\label{firstpage}
\pagerange{\pageref{firstpage}--\pageref{lastpage}}
\maketitle

%%%%%%%%%%%%%%%%%%%%%%%%%%%%%%%%%%%%%%%%%%%%%%%%%%%%%%%%%%%%%%%%%%%%%%%%%%%%%%%%%%%
% Abstract of the paper
%%%%%%%%%%%%%%%%%%%%%%%%%%%%%%%%%%%%%%%%%%%%%%%%%%%%%%%%%%%%%%%%%%%%%%%%%%%%%%%%%%%
 \begin{abstract}
The direction of the spin vectors of disk galaxies change over time.
We present the tilting rate of a sample of galaxies in the NIHAO suite of cosmological hydrodynamical simulations.
Galaxies in our sample have been selected to be isolated and to have well determined spins.
We compare the tilting rates to the predicted observing limit of {\it Gaia}, finding that our entire sample lies above the limit, in agreement with previous work. To test the role of dark matter and of gas we compare the weighted Pearson's correlation coefficients between the tilting rates and various properties.
We find no correlation between the dark halo's tilting rate, shape, or misalignment with respect to the disc, and the tilting rate of the stellar disc.
Therefore, we argue that, in the presence of gas, the dark halo plays a negligible role in the tilting of the stellar disc.
On the other hand, we find a strong correlation between the tilting rate of the stellar disc and the misalignment of the cold gas warp.  
Adding the stellar mass fraction improves the correlation, while none of the dark matter's properties together with the cold gas misalignment improves the correlation to any significant extent.
This implies that the gas cooling onto the disc is the principal driver of disc tilting.
\end{abstract}


%%%%%%%%%%%%%%%%%%%%%%%%%%%%%%%%%%%%%%%%%%%%%%%%%%%%%%%%%%%%%%%%%%%%%%%%%%%%%%%%%%%
% Introduction
%%%%%%%%%%%%%%%%%%%%%%%%%%%%%%%%%%%%%%%%%%%%%%%%%%%%%%%%%%%%%%%%%%%%%%%%%%%%%%%%%%%
\section{Introduction}
\label{sec:intro}
Over the course of its five-year mission, {\it Gaia} will provide accurate distances and proper motions for roughly 
one per cent of the stars in the Milky Way (MW) \citep{perryman2001}, as well as positions of some $500,000$ quasars, with 
redshifts between $z = 1.5{\text -}2$ \citep{perryman2001,claeskens2006,lindegren2008}.
These quasars will provide an inertial reference frame accurate to $0.2{\text -} 0.5 \mu$as yr$^{-1}$.
As a result, \cite{perryman2014} predicted an accuracy better than $1 \mu $as yr$^{-1}$ ($0.27 \degrees$ Gyr$^{-1}$) should be 
achieved in all inertial spin components of the {\it Gaia} reference frame.
\citet{BinneyMay86} were the first to propose that discs slew as misaligned angular momentum is accreted by a galaxy.
\citet{OstrikerBinney89} attributed the formation of warps to disc slewing.
Moreover dark haloes also tumble, which may also drive tilting of the disc.  If the tilting rates of the stellar discs are similar to those of dark haloes \citep{bailinsteinmetz2004,bryancress2007}, such a tilt would be detectable in the MW.
\cite{earp2017} (hereafter, E17) presented the tilting rates for galaxies comparable to the Milky Way ($9 \times 10^{11} \Msun \le M_{200} \le 1.2 \times 10^{12} \Msun $) in a $\Lambda$ Cold Dark Matter cosmological hydrodynamical simulation.
E17 measured the tilting rates between $z=0.3$ and $0$, excluding mergers, finding that all galaxies exhibited significant tilting, at an average rate of $4.9\degrees \pm 2.7\degrees$Gyr$^{-1}$, well above {\it Gaia's} predicted detection limit.

Various processes can cause discs to tilt.
Minor mergers and small-scale interactions can result in a change in angular momentum \citep{ostrikertremaine1975,toomre1981,huangcarlberg1997,sellwood1998,benson2004,read2008,kazantzidis2009,frings2017}.
As satellites fall into the disc they tilt towards alignment with the disc due to the transfer of angular momentum \citep{huangcarlberg1997,read2008}.
The most massive interaction of the MW is with the Large Magellanic Cloud (LMC), with mass estimates as high as $M_\mathrm{LMC} \sim 2 \times 10^{11}\Msun$ \citep{kallivayalil2013, gomez2015, penarrubia2015}, corresponding to $20$ per cent of the mass of the MW.
If the LMC is this massive it would put it above the upper mass estimate for the initial mass of the Sagittarius dwarf galaxy \citep{jiang2000}, meaning it is the most massive interaction in some time.
Because of the uncertainty in the LMC's mass, its importance on the tilting of the MW's disc cannot yet be determined.

Sustaining star formation in galaxies requires ongoing gas accretion onto the disc.
As gas falls into the dark matter's potential well it is shock heated to the halo virial temperature $T_{\rm vir} = 10^6 (\nu_{\rm circ} / 167\text{km s}^{-1})^2$K, forming a hot gas corona \citep{spitzer1956, reesostriker1977, silk1977, white1978, savageboer1979, white1991, dahlem1997,  wang2001,fukugita2006}.
This hot gas cools and eventually settles into the disc \citep{fall1980,brook2004,keres2005,robertson2006,brooks2009}.
Gas cooling from the corona contributes angular momentum to the disc.
Cosmological simulations have shown that the coronae of MW progenitor galaxies form early and thereafter their angular momentum evolves differently from that of the dark matter \citep[e.g.][]{obreja2019,jiang2019}. The angular momentum of the corona is usually misaligned with that of their stellar disc \citep{vandenbosch2002,roskar2010,velliscig2015,stevens2017}.
This addition of misaligned angular momentum results in the disc tilting.
\citet{debattista2015} (hereafter, D15) showed that under these circumstances, the orientation of the disc's angular momentum is determined by a balance between the torques from the triaxial dark halo, and the net inflow of angular momentum via cooling gas.
As a result, star-forming galaxies where gas is continually cooling onto the disc are generally misaligned with the principal axes of their dark haloes, as has been found in large surveys \citep{sales2004, brainerd2005, agustsson2006, yang2006,azzaro2007,faltenbacher2007,wang2008a,wang2010, nierenberg2011, li2013}. 

\cite{dubinski1992} measured the tilting rates of dark haloes in the mass range $(1-2) \times 10^{12} \Msun$; he found that they rotated uniformly with rotation rates in the range $6 \degrees - 96\degrees $Gyr$^{-1}$.
Similarly \cite{bailinsteinmetz2004} found smooth figure rotation with an average tilting rate of $6.2\degrees$Gyr$^{-1}$.
\cite{bryancress2007} found that $63$ haloes exhibited an average pattern speed of $13.8 \degrees \; h \;$Gyr$^{-1}$.
To investigate coupling between the orientation of discs and dark haloes, \cite{yurinspringel2015} inserted live stellar discs into eight, MW-sized, high-resolution dark haloes from the {\sc aquarius} simulation.
They found tilting rates of $5\degrees - 6 \degrees$Gyr$^{-1}$, comparable to the halo tilting rates measured in pure {\it N}-body simulations.

\cite{bettfrenk2012} examined the consequences of minor mergers and flybys on the spin of dark haloes.
They measured the angular momentum of haloes with mass $12.0 \le \log_{10} (M/\Msun) h^{-1} \le 12.5$ at $z=0$. 
They found that such events only caused small changes to the angular momentum of the entire halo.
However, the inner halo ($R < 0.25R_{200}$), which \citet{Binneyetal98} show is very tightly coupled gravitationally to the stellar disc, was not so stationary, with 47 per cent experiencing a change in their angular momentum orientation of at least $45 \degrees$ during their lifetimes.

Therefore, there is a good theoretical basis to expect the MW's disc to be tilting, with several possible driving mechanisms.
In this paper, we investigate the role of the dark matter and of gas on the tilting rates of stellar discs from the NIHAO suite of cosmological hydrodynamical zoom-in simulations.
The paper is organized as follows.
Section \ref{sec:simulation} presents the suite of simulations used in this paper.
Section \ref{sec:samples} details the selection of a subsample of galaxies that have well-determined tilting rates.
Section \ref{sec:drivers} compares the tilting rates to various possible predictors.
Section \ref{sec:conclusions} presents our conclusions.

%%%%%%%%%%%%%%%%%%%%%%%%%%%%%%%%%%%%%%%%%%%%%%%%%%%%%%%%%%%%%%%%%%%%%%%%%%%%%%%%%%%
% Numerical simulations
%%%%%%%%%%%%%%%%%%%%%%%%%%%%%%%%%%%%%%%%%%%%%%%%%%%%%%%%%%%%%%%%%%%%%%%%%%%%%%%%%%%
\section{Numerical Simulation}
\label{sec:simulation}

For this paper we used the NIHAO (Numerical Investigation of Hundred Astrophysical Objects)\footnote{'nihao' is the Chinese word for {\it 'hello'}}
simulations suite. NIHAO is a sample of $\approx 100$ hydrodynamical cosmological zoom-in simulations performed using the SPH code {\sc gasoline2} \citep{wadsley2017}. 
The code includes gas heating via ultraviolet (UV) heating and ionization and cooling due to hydrogen, helium and metals \citep{shen2010}. The star formation and feedback modelling follows what was used in the MaGICC simulations \citep{stinson2013}, adopting a threshold for star formation of $n_{th} > 10.3$ cm$^{−3}$. Stars can feed energy back into the interstellar medium via SN feedback \citep{stinson2006}  and via ionizing radiation from massive stars (early stellar feedback) before they turn into SN \citep{stinson2013}. We refer the reader to \cite{wang2015} for a more detailed description of the code and the simulations.

\begin{figure}
  \begin{center}
    \includegraphics[width=\columnwidth]{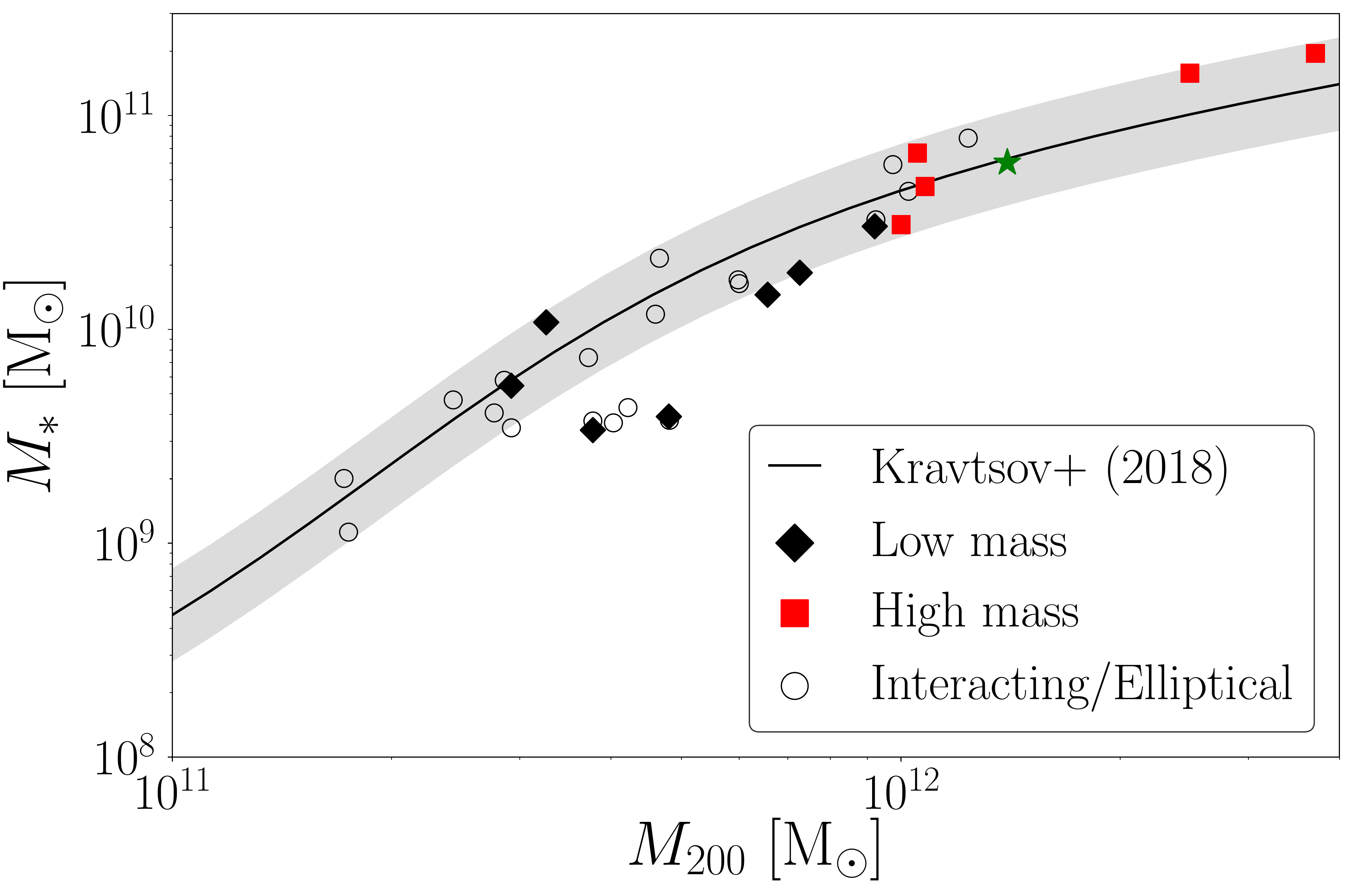}
    \caption[Abundance matching for NIHAO galaxies.]{Halo mass versus stellar mass for the excluded galaxies (circles), low mass sub-sample galaxies (black diamonds) and high mass sub-sample galaxies (red squares). All samples trace the observed abundance matching relation of \citet{kravtsov2018} (black line with the grey shaded region showing the $1\sigma$ scatter). The green star denotes the MW, assuming a halo mass of $M_{200} \sim 1.1 \times 10^{12} \Msun$ and a stellar mass of $M_{*} = 5 \times 10^{10} \Msun$ \citep{blandhawthorn2016}.}
    \label{fig:abundance_matching_halo_mass}
  \end{center}
\end{figure}

The NIHAO simulations are the largest set of cosmological zoom-in simulations covering the halo mass range $10^{10}$ - $10^{12} \Msun$,
they combine very high spatial and mass resolution with a statistical sample of haloes. 
NIHAO has proven very successful in reproducing several key properties of observed galaxies including their
cold gas masses and sizes \citep{stinson2015,maccio2016}, the stellar and baryonic Tully-Fisher relations \citep{dutton2017} and stellar disc kinematics \citep{obreja2016}.
A key property of the NIHAO galaxies is that they lie on the stellar mass-halo mass relation
across their full mass range. This is shown for the galaxies used in this work in Figure  \ref{fig:abundance_matching_halo_mass}, 
where the virial mass versus the stellar mass within 10 per cent of $R_{200}$ for NIHAO galaxies is compared to the observed relationship from \citet{kravtsov2018}, derived using halo abundance matching (black line).
The green star on Figure \ref{fig:abundance_matching_halo_mass} denotes the observed MW values, with virial mass $M_{200} \sim 1.1 \times 10^{12} \Msun$ and stellar mass $M_{*} \sim 5 \times 10^{10} \Msun$ \citep{blandhawthorn2016}.
Having galaxies with realistic stellar content at all masses allows us to extend the analysis performed in E17, and hence to consider the tilting rates of lower mass galaxies.

%%%%%%%%%%%%%%%%%%%%%%%%%%%%%%%%%%%%%%%%%%%%%%%%%%%%%%%%%%%%%%%%%%%%%%%%%%%%%%%%%%%
% Sample selection
%%%%%%%%%%%%%%%%%%%%%%%%%%%%%%%%%%%%%%%%%%%%%%%%%%%%%%%%%%%%%%%%%%%%%%%%%%%%%%%%%%%
\section{Sample selection}
\label{sec:samples}

\subsection{Tilting rates}
To measure the angular momentum of our galaxies, it is vital that their centres are correctly determined.
We use two different methods to determine the centre of each galaxy: the centre of mass within a shrinking sphere, and the lowest potential dark matter particle.
The shrinking sphere method follows the iterative technique of \cite{power2003}: at each step, the centre of the sphere is determined as the centre of mass of the previous step, and the radius of the sphere is reduced by 2.5 per cent.
This is iterated down to the smallest sphere containing at least 100 particles, at which point the centre of mass of this final sphere is returned.
For the second technique, we simply look for the dark matter particle with the lowest potential energy and use its position as the centre.
As in E17, we find that the lowest potential dark matter particle gives more reliable centres for the sample, due to some galaxies having high-density substructure away from the true centre.
After removing galaxies where the centres were still poorly determined, the sample is reduced to 85 galaxies from 91.
We are predominantly interested in MW mass galaxies, therefore, we impose a lower limit on the stellar mass of $M_* > 10^9 \Msun$, leaving us with 30 galaxies.

The angular momentum of the stellar disc is calculated using only stars with ages $\tau < 1$ Gyr within a range of radii, from 0.5 per cent of $R_{200}$, out to 10 per cent of $R_{200}$, in 0.5 per cent increments.
We adopt the angular momentum of the young stars within 5 per cent of $R_{200}$ ($\sim 10$ kpc for a MW sized galaxy) for the stellar disc.
This is motived by warps often being present at radii, $R > 0.05 R_{200}$.
We determine the tilting rates by measuring the angle, $\Delta \theta$, between the angular momentum vectors of the stellar disc at $z=0.3$ and at $0$, then dividing by the time difference ($\sim 3.7$ Gyr).

We have verified, by visual inspection, that the evolution of the tilt is uniform over 20 steps between $z=0.3$ and $z=0$, indicating that the tilting is coherent. Thus the integrated tilt is not a spurious effect of random noise.  We will present examples of this evolution in a followup paper.

%%%%%%%%%%%%%%%%%%%%%%%%%%%%%%%%%%%%%%%%%%%%%%%%%%%%%%%%%%%%%%%%%%%%%%%%%%%%%%%%%%%

\subsection{Isolation criterion}
\begin{figure}
  \begin{center}
    \includegraphics[width=\columnwidth]{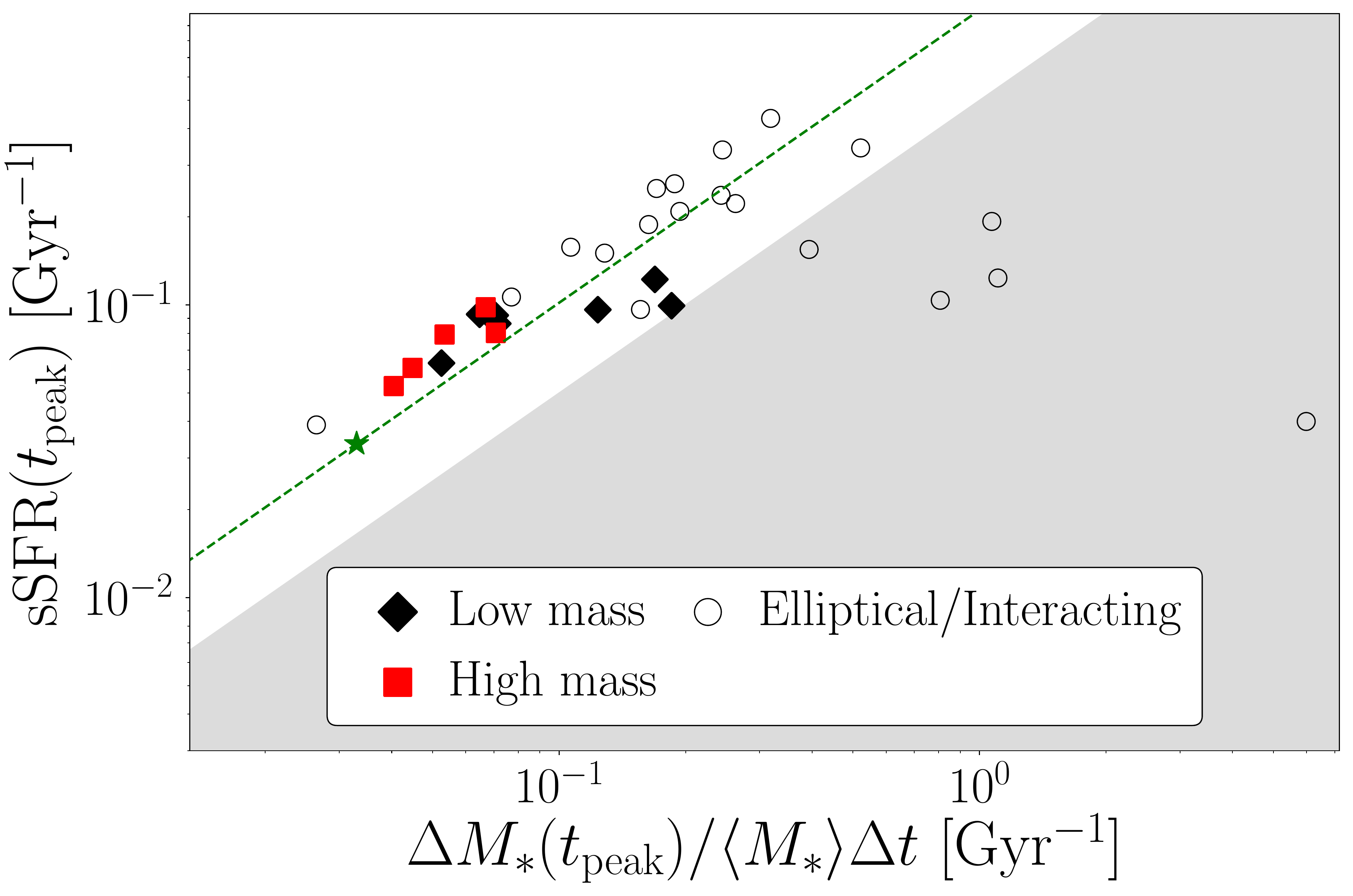}
    \caption[Maximum stellar mass change versus specific star formation rate, with samples A and B defined.]{Maximum stellar mass change versus specific star formation rate over the same time. The (black) diamonds show the low mass galaxies in our subsample, the (red) squares show the high mass galaxies and the open circles show excluded galaxies, which are ellipticals or are strongly interacting. The (green) star indicates the Galaxy, derived from values of the sSFR and supernovae rate from \citet{blandhawthorn2016} and \citet{diehl2006}, respectively. The (green) dashed line applies an offset to the one-to-one relation by assuming a MW supernovae rate from \citet{diehl2006}. The grey shaded region highlights galaxies that gained most of their stellar mass directly from accretion.}
    \label{fig:secondsamplecut}
  \end{center}
\end{figure}

Following E17, we are primarily interested in galaxies evolving in relative isolation.
In such galaxies, the change in stellar mass should be driven primarily by in-situ star formation.
Therefore, we compare the maximum fractional change in stellar mass $\Delta M_*(t_{\rm peak}) / (\langle M_* \rangle \Delta t)$ to the specific star formation rate (sSFR), given by the ratio of the star formation rate to the stellar disc mass, at the same time step, $t_{\rm peak}$.
We further refine our high mass galaxies by excluding any galaxy that has gained more than 50 per cent of its stellar mass via accretion at any time step; leaving us with 26 galaxies.
We then divide these 26 galaxies into two mass subsamples, {\it high mass} (6 galaxies) and {\it low mass} (20 galaxies), with $M_{200} = 10^{12} \Msun$ being the separator.
Figure \ref{fig:secondsamplecut} shows the resulting distribution in the space of mass growth versus sSFR.
Galaxies in the shaded region will have gained a majority of stellar mass through accretion and are excluded.
The (black) diamonds denote the low mass galaxies and the (red) squares show the high mass galaxies. We estimate the mass lost due to supernova feedback by multiplying the rate for  type-II supernovae in the MW ($1.9 \pm 1.1$ per century \citet{diehl2006}) by the integral of the Kroupa IMF \citep{kroupa2002} for high mass stars ($M_* = 8 - 40 \Msun$) derived by the integral for all mass.
Combining this feedback rate with the Galaxy's sSFR rate obtained from the current star formation rate
$1.65 \pm 0.19 \Msun$ yr$^{-1}$ and the current stellar mass, $M_{*} \sim 5 \times 10^{10} \Msun$ \citep{blandhawthorn2016}, we estimate the stellar mass change of the MW.
This implies that $\sim 30$ per cent of the mass gained from star formation is lost due to supernovae feedback. The (green) dashed line in Fig. \ref{fig:secondsamplecut} applies this offset to the one-to-one relationship.

% Sample plot

%%%%%%%%%%%%%%%%%%%%%%%%%%%%%%%%%%%%%%%%%%%%%%%%%%%%%%%%%%%%%%%%%%%%%%%%%%%%%%%%%%%

\subsection{Error measurements}
The spin direction errors were obtained using two different methods. 
First, we calculate the angular momentum vectors of the young stellar disc within 2, 3, 4 and 5 per cent of $R_{200}$.
We then measure the angle between each of these four vectors, $\phi$, and their sum.
The mean of these values is then taken as our error,
$$\epsilon_R(t) =  \Bigg \langle \phi \Bigg({{\bm {L}}(R,t),\sum_R {\bm L}}(R,t) \Bigg), \;\;\; R=2,3,4,5 \Bigg \rangle.$$
We take the root mean squared sum of $\epsilon_R(t)$ at $z=0.3$ and $0$, resulting in our first error $\epsilon_R$.  We use only stars with ages $\tau < 1$ Gyr for this calculation, as these are the least contaminated by the bulge and stellar halo; because of this we must ensure that the young stars are closely aligned to the rest of the disc.
Therefore, we employ a second method by calculating the angular momentum vectors of star particles with ages in 2 Gyr bins from 0 to 8 Gyr.
As before, we measure the angle between each of these four vectors, $\phi$, and their sum.
We then take the mean value,
$$\epsilon_\tau(t) =  \Bigg \langle \phi \Bigg({{\bm {L}}(\tau,t),\sum_\tau {\bm L}}(\tau,t) \Bigg), \;\;\; \tau=\tau_{0,2},\tau_{2,4},\tau_{4,6},\tau_{6,8} \Bigg \rangle,$$
where $\tau_{n,m}$ are the stars with ages between $n$ and $m$ Gyr.
We take the root mean squared sum of $\epsilon_\tau(t)$ at $z=0.3$ and $0$, resulting in $\epsilon_\tau$.
We construct our final subsample of galaxies by imposing an upper limit on both $\epsilon_{R}$ and $\epsilon_{\tau}$ of $\epsilon \leq 5\degrees$.
The upper limit on $\epsilon_{R}$ reduces our subsample down to 21 galaxies and the upper limit on $\epsilon_{\tau}$ reduces our subsample down to 12 galaxies (7 low mass and 5 high mass, which we refer to as the high mass and low mass samples, respectively, hereafter) that have well-determined tilting rates. We have verified by visual inspection that the galaxies removed by these cuts are poor analogues of the Milky Way, generally because they are either elliptical or interacting.

%%%%%%%%%%%%%%%%%%%%%%%%%%%%%%%%%%%%%%%%%%%%%%%%%%%%%%%%%%%%%%%%%%%%%%%%%%%%%%%%%%%
% Possible drivers
%%%%%%%%%%%%%%%%%%%%%%%%%%%%%%%%%%%%%%%%%%%%%%%%%%%%%%%%%%%%%%%%%%%%%%%%%%%%%%%%%%%
\section{Drivers of tilting}
\label{sec:drivers}
We find error-weighted averages for the tilting rate of the entire low$+$high mass subsample of $3.8 \pm 2.3 \degrees$ Gyr$^{-1}$ and $3.6 \pm 2.4 \degrees$ Gyr$^{-1}$ for $\epsilon_R$ and $\epsilon_\tau$, respectively.
For just the high mass galaxies we find error-weighted averages of $3.8 \pm 2.8 \degrees$ Gyr$^{-1}$ and $3.7 \pm 2.7 \degrees$ Gyr$^{-1}$ for $\epsilon_R$ and $\epsilon_\tau$, respectively.

\subsection{Pearson's correlation coefficient}
Throughout this section we report the error-weighted Pearson correlation coefficient, $\text{p}(x,y,\omega)$ between the tilting rate of the stellar disc and each possible predictor.
The weights are defined as $\omega = \epsilon^{-2}$, where $\epsilon$ is the error on the tilting rate.
With this set of weights we determine error-weighted covariance as
$$\text{cov}(x,y,\omega) = \frac{\Sigma_i \omega_i (x_i - \langle x, \omega \rangle))(y_i - \langle y, \omega \rangle)}{\Sigma_i \omega_i},$$
where $\langle x, \omega \rangle$ denotes the weighted mean of $x$. The error weighted Pearson correlation coefficient is given by
$$\text{p} = \frac{\text{cov}(x,y,\omega)}{\sqrt{\text{cov}(x,x,\omega) \text{cov}(y,y,\omega)}}.$$
We report p values for the entire low mass$+$high mass subsample, due to the small number of galaxies.
Correlation coefficients of $|{\rm p}| < 0.4$ will be treated as null results, values between $0.4 < |{\rm p}| < 0.7$ will be referred to as weak correlations and values of $|{\rm p}| > 0.7$ will be referred to as strong correlations.
Table \ref{tab:p_values} provides a full list of all correlation coefficients calculated using both error methods, $\epsilon_R$ and $\epsilon_\tau$, with strong correlations indicated in bold; however, in the text we will refer only to p values calculated using $\epsilon_R$.

\begin{table}
 \centering
  \begin{tabular}{llll}
   \hline
   Baryonic predictor & units & p($\epsilon_{r}$) & p($\epsilon_{\tau}$) \\
   \hline
   $M_*$ & M$_\odot$ & -0.16 & -0.19 \\
   $\Delta M_{\rm{max},*}/ \langle M_* \rangle \Delta t$ & Gyr$^{-1}$ & -0.28 & -0.11 \\
   $\Delta M_{\rm{max},b}/ \langle M_{\rm b} \rangle \Delta t$ & Gyr$^{-1}$ & 0.05 & 0.14 \\
   $\Delta \theta_* / \Delta t$ & $^\circ$ Gyr$^{-1}$ & -0.00 & -0.01 \\
   $\theta (*, \rm{corona})$ & $^\circ$ & 0.03 & -0.02 \\
   $\Delta \theta_{\rm{corona}} / \Delta t$ & $^\circ$ Gyr$^{-1}$ & 0.12 & 0.14 \\
   $\theta(*,\rm{cold \; gas})$ ($z = 0.3$) & $^\circ$ & {\bf 0.82} & {\bf 0.82} \\
   $\theta(*,\rm{cold \; gas})$ ($z = 0.15$) & $^\circ$ & {\bf 0.86} & {\bf 0.89} \\
   $\theta(*,\rm{cold \; gas})$ ($z = 0$) & $^\circ$ & 0.62 & 0.63 \\
   log$_{10}$ sSFR & Gyr$^{-1}$ & 0.08 & 0.25 \\
   log$_{10}$ $\langle $sSFR$ \rangle$ & Gyr$^{-1}$ & -0.17 & -0.01 \\
   log$_{10}$ sSFR($z=0.3$) & Gyr$^{-1}$ & -0.42 & -0.34 \\
   log$_{10}$ max(sSFR) & Gyr$^{-1}$ & -0.39 & -0.35 \\
   $M_* / M_{*,200}$ & & 0.18 & 0.09 \\
   $M_* / M_{\rm{gas,200}}$ & & 0.26 & 0.18 \\
   $M_{\rm{gas,disc}} / M_*$ & & -0.19 & -0.09 \\
   $M_{\rm{gas,disc}} / M_{\rm{gas,total}}$ & & -0.06 & 0.07 \\
   $|L_*|(t_0)$ & $\Msun$ kpc km s$^{-1}$ & -0.44 & -0.36 \\
   $\Delta |L_*| / \langle |L_*| \rangle$ & & 0.63 & {\bf 0.70} \\
   $|L_{\rm cold}|(t_0)$ & $\Msun$ kpc km s$^{-1}$ & 0.11 & 0.23 \\
   $\Delta |L_{\rm cold}| / \langle |L_{\rm cold}| \rangle$ & & -0.16 & -0.10 \\
   \hline
   Dark matter predictor & units & p($\epsilon_{r}$) & p($\epsilon_{\tau}$) \\
   \hline
   $b/a$ & & 0.11 & 0.15 \\
   $c/a$ & & 0.08 & 0.07 \\
   $T$ & & -0.26 & -0.37 \\
   $|$cos $\theta (*, \rm{major})|$ & $^\circ$ & -0.35 & -0.27 \\
   $|$cos $\theta (*, \rm{intermediate})|$ & $^\circ$ & -0.23 & -0.52 \\
   $|$cos $\theta (*, \rm{minor})|$ & $^\circ$ & -0.27 & -0.18 \\
   $\rho / \rho_{\rm crit}$(R < 4 Mpc) & & -0.16 & -0.08 \\
   $\rho / \rho_{\rm crit}$(R < 6 Mpc) & & -0.05 & 0.21 \\
   $\rho / \rho_{\rm crit}$(R < 8 Mpc) & & 0.38 & 0.50 \\
   $\Delta \theta_{\rm DM}(R < 0.05 R_{200}) / \Delta t$ & $^\circ$ Gyr$^{-1}$ & {\bf 0.89} & {\bf 0.95} \\
   $\Delta \theta_{\rm DM}(R < 0.1 R_{200})  / \Delta t$ & $^\circ$ Gyr$^{-1}$ & {\bf 0.84} & {\bf 0.87} \\
   $\Delta \theta_{\rm DM}(R < R_{200})  / \Delta t$ & $^\circ$ Gyr$^{-1}$ & 0.34 & 0.64 \\
   $\theta(\rm DM, *, R < 0.1 R_{200})$ & $^\circ$ & -0.17 & -0.36 \\
   $\theta(\rm DM, *, R < 0.1 R_{200})$ & $^\circ$ & -0.10 & -0.21 \\
   $\theta(\rm DM, *, R < R_{200})$ & $^\circ$ & -0.23 & -0.24 \\
   \hline
   \end{tabular}
  \caption{All error-weighted Pearson's correlation coefficients reported in this paper, between the predictor listed and the tilting rate of the stellar disc ($\Delta \theta_* / \Delta t$). Bold p values highlight correlations with ${\rm p} > 0.7$. When logarithmic values are given or the values are plotted in log-space the correlation coefficient is calculated after taking the logarithm and assuming a linear relationship.}
 \label{tab:p_values}
\end{table}

%%%%%%%%%%%%%%%%%%%%%%%%%%%%%%%%%%%%%%%%%%%%%%%%%%%%%%%%%%%%%%%%%%%%%%%%%%%%%%%%%%%
\subsection{The minimal role of dark matter}
In this section we compare the properties of the dark halo to the tilting rates of the stellar discs they harbour, to investigate the extent of the halo's influence. 

\begin{figure*}
  \begin{center}
    \includegraphics[width=\textwidth]{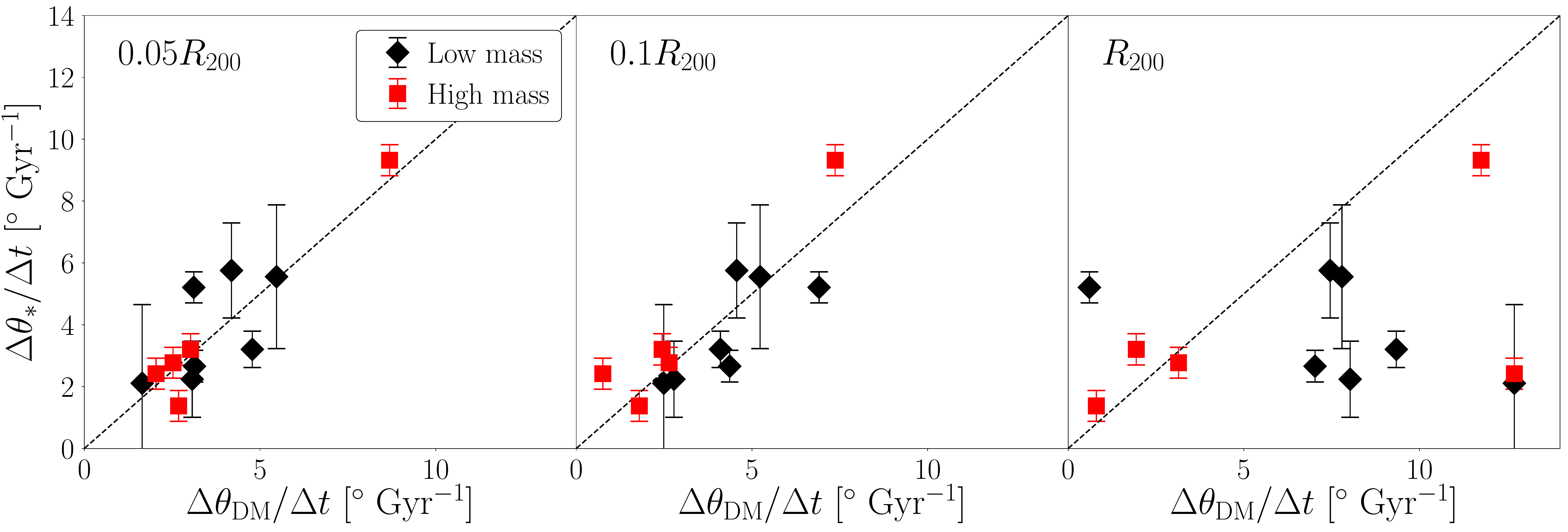}
    \caption[Tilting rates of both the stellar and dark matter components.]{Tilting rate of the stellar disc ($R < 0.05 R_{200}$) and the dark matter for low mass galaxies (black diamonds) and high mass galaxies (red squares), where the angular momentum of the dark matter is calculated within $0.05 R_{200}$ (left), $0.1 R_{200}$ (middle) and $R_{200}$ (right). The (black) dashed line represents the one-to-one relationship. Both the left and middle panels show a strong correlation, which vanishes in the right panel.}
    \label{fig:darkmomdthetadt}
  \end{center}
\end{figure*}

Figure \ref{fig:darkmomdthetadt} presents the tilting rates of the stellar disc and the dark matter's angular momentum. 
The stellar disc's angular momentum is measured within $0.05 R_{200}$ for all three panels, whereas, the dark matter's angular momentum is measured within $0.05 R_{200}$ (left), $0.1 R_{200}$ (middle) and $R_{200}$ (right).
The distribution roughly follows the one-to-one relationship (black dashed line) within $R < 0.1 R_{200}$, with a very strong correlation for both the left and middle panels.
For $R_{200}$ (right panel), this correlation vanishes, suggesting that the tilting of the entire dark halo is not related to that of the disc.
Instead, just the inner part of the halo, which is directly affected by the disc, follows the disc closely.

\begin{figure*}
  \begin{center}
    \includegraphics[width=\textwidth]{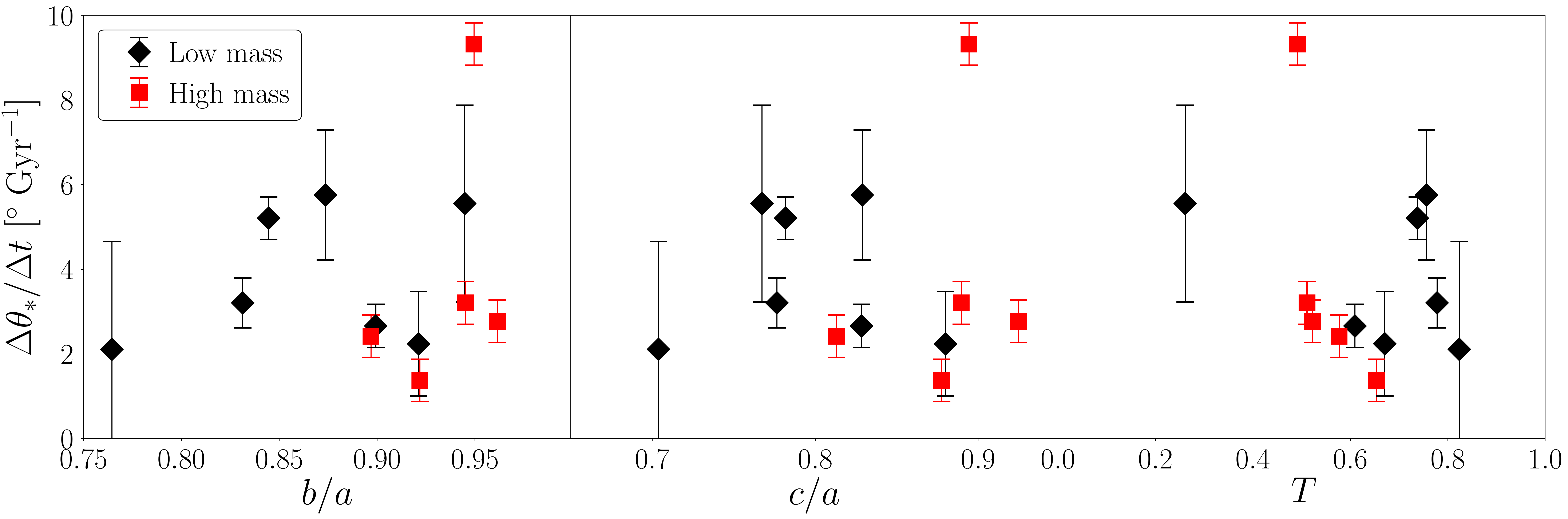}
    \caption[Tilting rate versus halo axis ratios and triaxiality.]{Intermediate to major axes ratio (left), minor to major axes ratio (middle) and triaxiality (right) versus tilting rate for both low mass galaxies (black diamonds) and high mass galaxies (red squares), at $z=0$. We find no correlation between the shape of the halo and the tilting rate of the stellar disc for $z=0$; a similar analysis at $z=0.3$ also finds no correlation.}
    \label{fig:haloshapedthetadt}
  \end{center}
\end{figure*}

As the shape of the halo becomes less spherical the torques imposed will increase, all other things being equal.
Therefore, if the dark halo is the primary driver of the tilting stellar disc, one might expect a correlation between the shape of the halo and the tilting rate.
To measure the shape of the halo, we define the principal axes of the dark halo with the condition that $a>b>c$ and, following \cite{kazantzidis2004}, the principal axis ratios are given by $s=b/a$ and $q=c/a$. 
Using an iterative process, the shape of the dark halo is calculated, starting with a spherical ellipsoid. 
The modified inertia tensor $I_{ij}$ is defined as:
$$I_{ij} = \sum_\alpha m_\alpha x_i^\alpha x_j^\alpha / r_\alpha^2$$
where $x_i^\alpha$ is the $i$th coordinate of the $\alpha$th particle and $r_\alpha$ is the elliptical radius defined as $r_\alpha^2 = x^2_\alpha + y^2_\alpha / s^2 + z_\alpha^2 /q^2$.
The eigenvalues of the modified inertia tensor are used as the new values of $s$ and $q$.
These iterations continue until the values of $s$ and $q$ converge to a fractional difference less than $10^{-2}$.
Following \cite{franx1991} we measure the triaxiality as 
$$T = \frac{1- (b/a)^2}{1 - (c/a)^2}.$$
$T$ values of $1$, $0$ and $\sim0.5$, correspond to prolate, oblate and triaxial spheroids, respectively. 
Figure \ref{fig:haloshapedthetadt} shows the tilting rate of the disc with respect to the resulting intermediate to major axis ratios ($b/a$, left), minor to major axes ratios ($c/a$, middle) and the triaxiality parameter ($T$, right), measured at $z=0$.  
Although previous authors have claimed that they are able to cause stellar discs to tilt purely by the gravitational dynamical interactions between the disc and halo, in the case of these hydrodynamical simulations we find no correlation between the shape of the halo and the tilting rate of the disc.
We have also verified this with the halo shape measured at $z=0.3$, again finding no correlations.

\begin{figure*}
  \begin{center}
    \includegraphics[width=\textwidth]{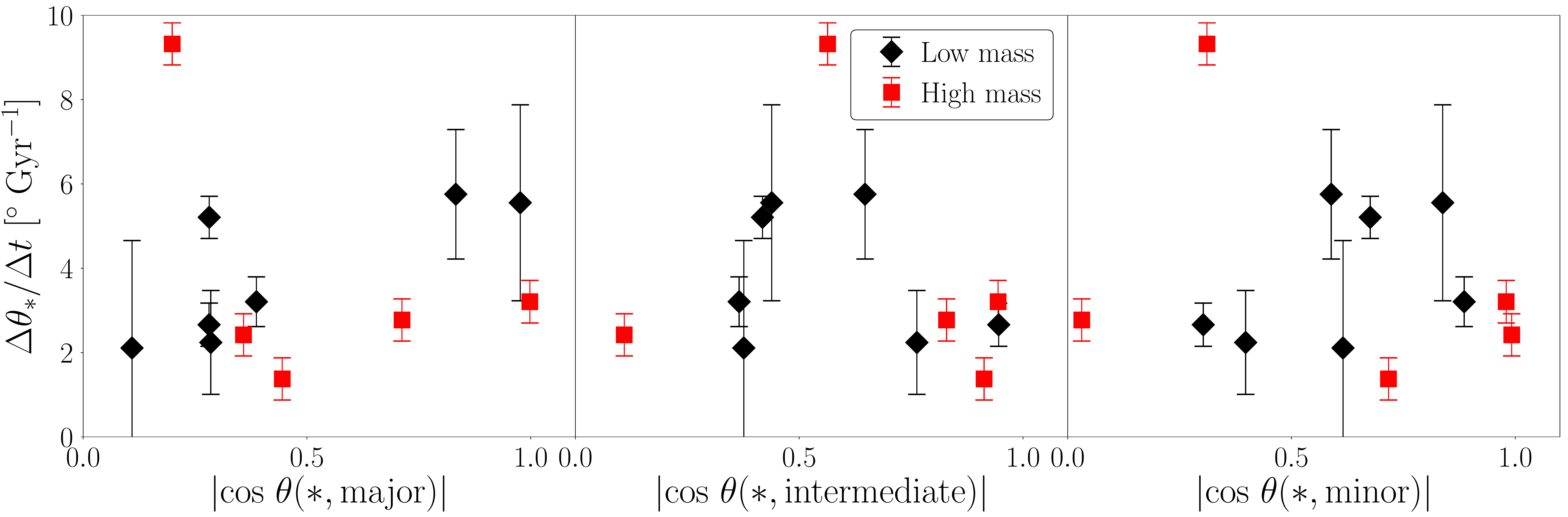}
    \caption[Tilting rate versus halo alignment.]{Offsets between the angular momentum of the stellar disc and the major- (left), intermediate- (middle) and minor-axes (right) versus the tilting rate of the stellar disc, at $z=0$. We find only a weak correlation between the alignment and the tilting rate for the intermediate axis; a similar analysis at $z=0.3$ finds no correlations.}
    \label{fig:haloalignmentdthetadt}
  \end{center}
\end{figure*}

D15 showed that red galaxies tended to be aligned such that their disc angular momentum was parallel to the minor axis of their parent dark halo, whereas blue galaxies tended to have random orientations.
To test if stellar discs tilt towards alignment with one of the principal axes of the dark halo, Figure \ref{fig:haloalignmentdthetadt} shows the distance between the angular momentum vector of the stellar disc and the major- (left), intermediate- (middle) and minor-axis (right); we find no correlations for any of the alignments, neither measured at $z=0.3$ nor at $z=0$.

\begin{figure*}
  \begin{center}
    \includegraphics[width=\textwidth]{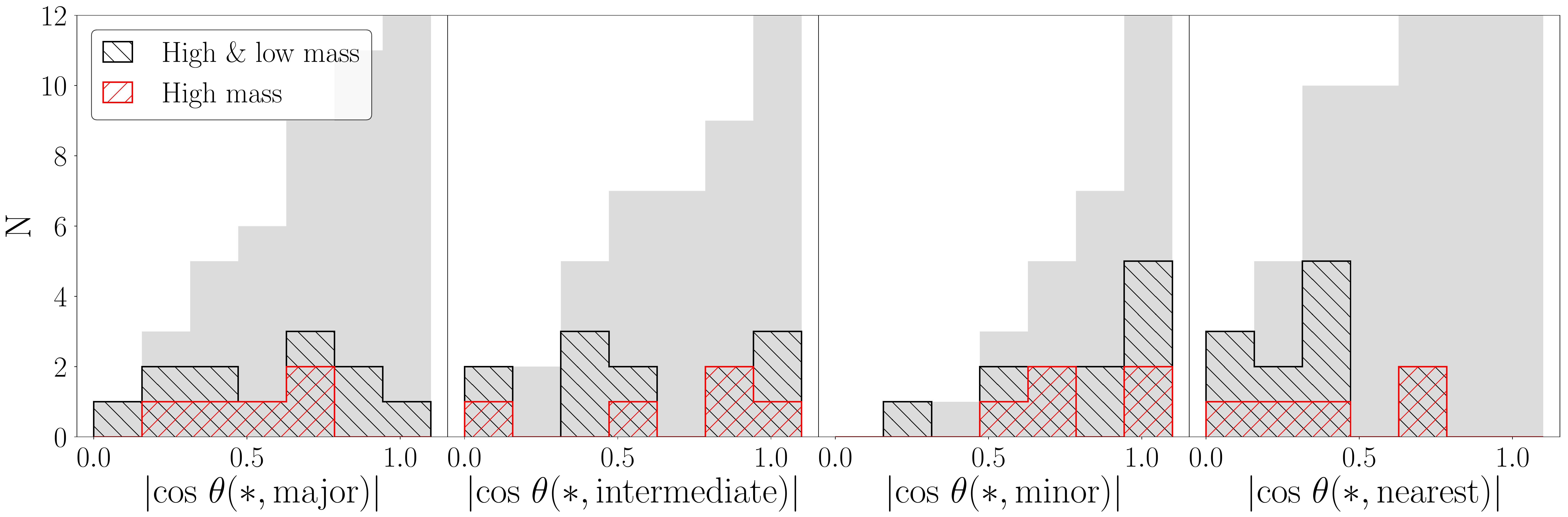}
    \caption[Distribution of halo alignments.]{Distribution of angular distances between the angular momentum of the stellar disc and the major- (left), intermediate- (middle left), minor- (middle right) and closest axis (right) of the dark halo. The black histogram shows high and low mass galaxies, and the red histogram shows only the high mass galaxies, the (grey) shaded region shows the cumulative distribution for high and low mass galaxies. We find both the high and low mass galaxies are predominantly star forming with no preferential alignment.}
    \label{fig:haloalignfull}
  \end{center}
\end{figure*}

Figure \ref{fig:haloalignfull} shows the distribution of alignments between stellar discs and their host haloes. 
The right panel shows that the majority of the galaxies are not closely aligned to any of the dark halo's principal axes.
Almost all the galaxies in the low mass subsample are star-forming blue galaxies, with sSFRs generally above the level of the MW.
This is in line with the result of D15 that blue galaxies generally have random alignments with respect to their dark halo.
We also considered the alignment between the hot ($T > 50000$K) gas corona and the dark halo's angular momenta, again finding no tendency to align.
Therefore, as the angular momentum of the corona is a product of the ongoing gas flow from the surrounding IGM and feedback, we argue that there is no preference for the angular momenta of the inflow to be aligned with that of the dark halo.

%%%%%%%%%%%%%%%%%%%%%%%%%%%%%%%%%%%%%%%%%%%%%%%%%%%%%%%%%%%%%%%%%%%%%%%%%%%%%%%%%%%

\subsection{Local cosmic overdensity}
\begin{figure*}
  \begin{center}
    \includegraphics[width=\textwidth]{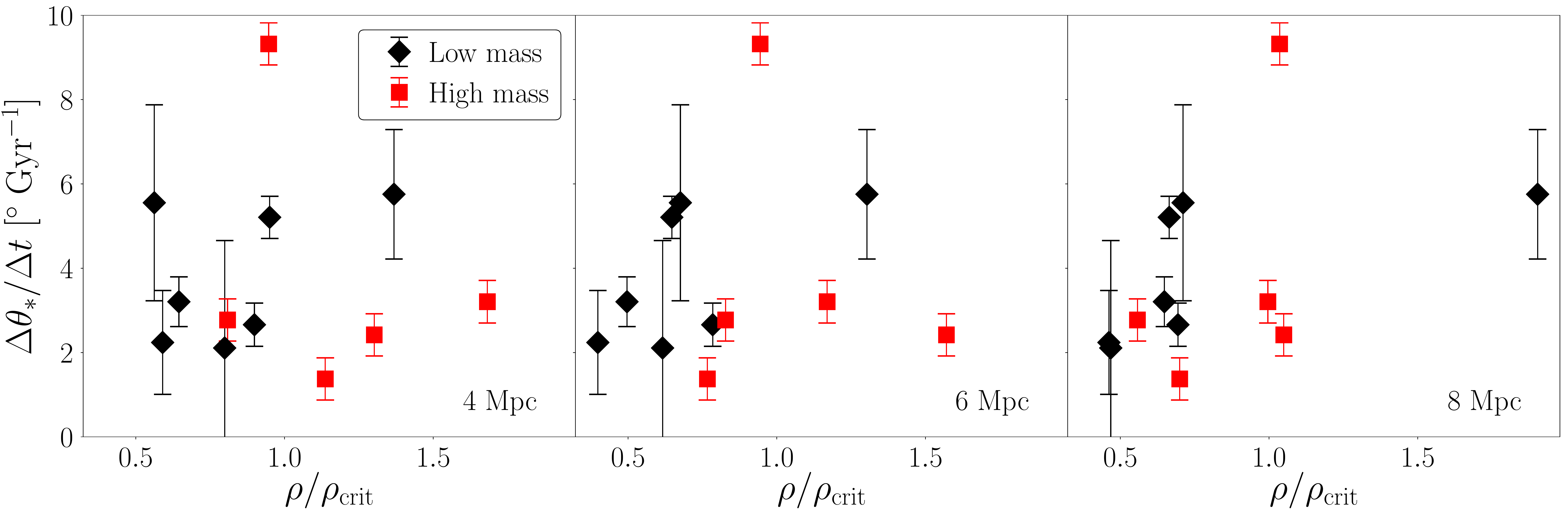}
    \caption[Tilting rate versus local density.]{Normalised local cosmic overdensity out to 4 Mpc (left), 6 Mpc (middle) and 8 Mpc (right), versus the tilting rate for low mass galaxies (black diamonds) and high mass galaxies (red squares), at $z=0$. We find no correlations for any radius at $z=0$; a similar analysis at $z=0.3$ also finds no correlations.}
    \label{fig:localdens2}
  \end{center}
\end{figure*}

Another possible driver of disc tilting is the proximity of high-density regions.
E17 found that galaxies in higher density regions tend to tilt at higher rates and this correlation strengthened for larger volumes.
Figure \ref{fig:localdens2} compares the tilting rate to the local density calculated within volumes of radii between 4 and 8 Mpc.
We find no correlation between the tilting rate of the stellar disc and the local cosmic overdensity out to any radius, at $z=0.3$ or at $z=0$.
This is in contrast to the results of E17. 
However, the galaxies in high-density regions in E17 were more likely to be in cluster/group environments, which might explain this apparent discrepancy, whereas the NIHAO sample was specifically designed for more isolated galaxies.
Indeed the range of cosmic overdensities in E17 spanned 0.7 to 10.1, with an average and median of 2.5 and 1.7, respectively, whereas for our NIHAO subsample the range is much smaller, 0.4 to 1.6, with average and median values of 0.85 and 0.78, respectively. 
In comparison, the local overdensity for the local group is roughly unity \citep{klypin2003,karachentsev2005}.

%%%%%%%%%%%%%%%%%%%%%%%%%%%%%%%%%%%%%%%%%%%%%%%%%%%%%%%%%%%%%%%%%%%%%%%%%%%%%%%%%%%

\subsection{The impact of gas}
\begin{figure}
  \begin{center}
    \includegraphics[width=\columnwidth]{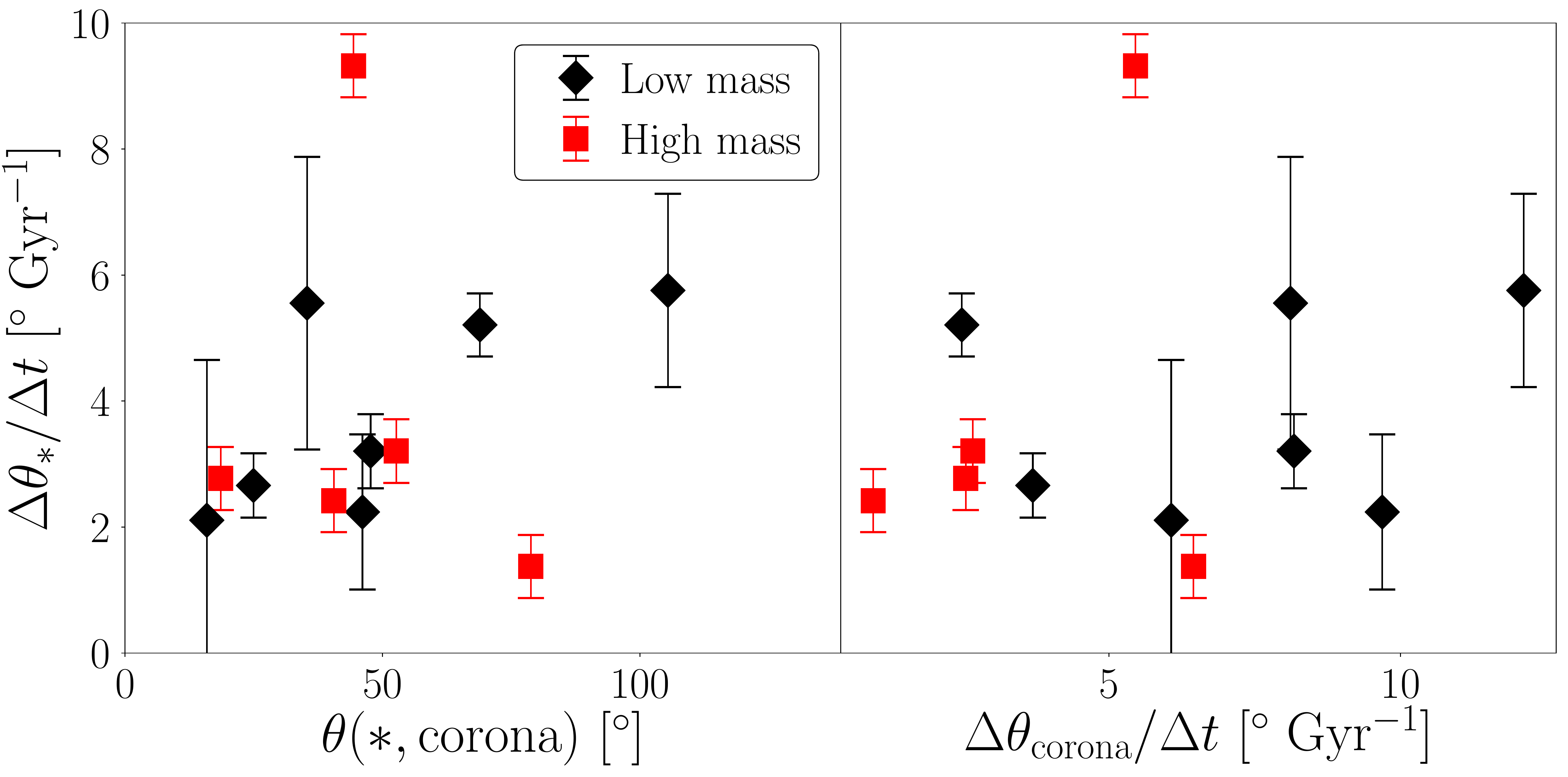}
    \caption[Tilting rate versus hot gas corona.]{Angular difference between the angular momentum vector of hot gas ($T > 50000$K) (left) and the tilting rate of the hot gas corona's angular momentum (right) versus the tilting rate of the disc, at $z=0$. We find no correlation in either panel; a similar analysis for gas parameters at $z=0.3$ also shows no correlation.}
    \label{fig:gasdiscdthetadt}
  \end{center}
\end{figure}

In hydrodynamical simulations, the stellar disc and hot gas corona are often misaligned \citep[e.g.][]{vandenbosch2002,velliscig2015,stevens2017}.
This misalignment results in gas cooling from the corona onto the disc with misaligned angular momentum, forming warps \citep{roskar2010}, and directly contributing misaligned angular momentum to the disc.
As in E17, we find that all our discs have angular momentum misaligned with that of their hot gas coronae.
Following E17, Figure \ref{fig:gasdiscdthetadt} compares both the angular misalignment between the hot ($T > 50000$K) gas corona and the stellar disc (left), as well as the tilting rate of the hot gas corona to the tilting rate of the disc (right).
We do not reproduce the weak correlation between the angular misalignment of the hot gas corona and disc found in E17.

Gas falling onto the hot gas corona from the intergalactic medium contributes misaligned angular momentum.
As a result, the net angular momentum of the hot gas corona tilts.
This gas then cools from the corona and contributes its misaligned angular momentum to the disc. 
Therefore, the two tilting rates may be correlated.
The right panel of Figure \ref{fig:gasdiscdthetadt} compares the two tilting rates; as in E17, we find no correlation.

\begin{figure*}
  \begin{center}
    \includegraphics[width=\textwidth]{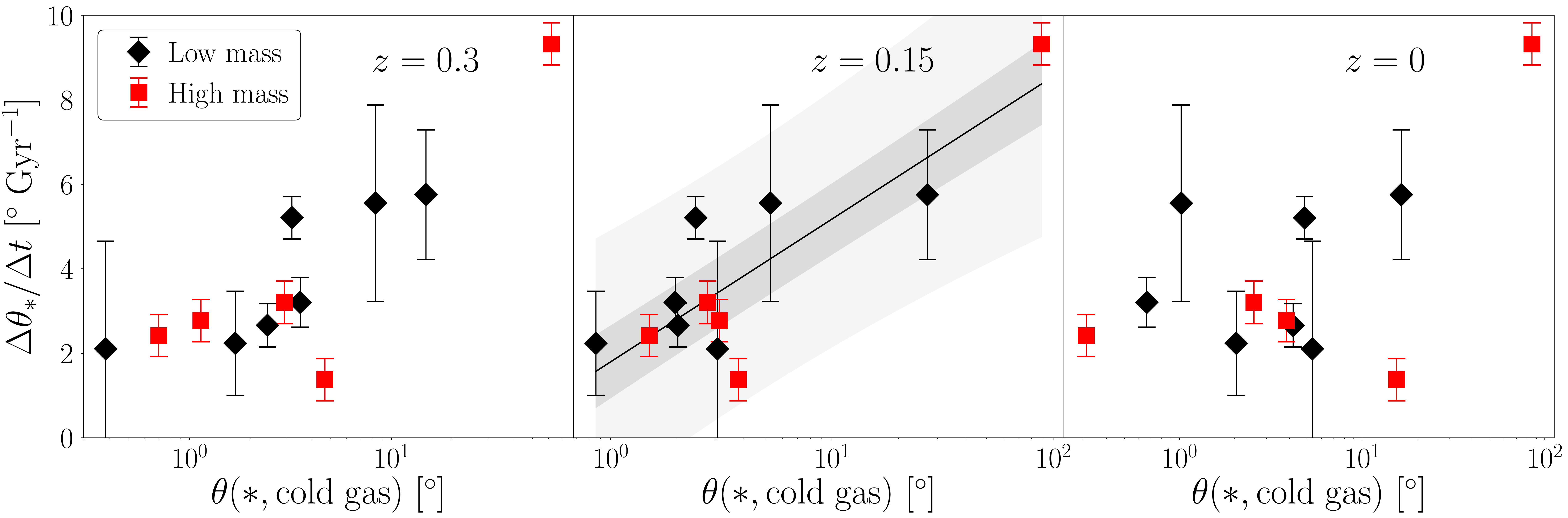}
    \caption[Tilting rate versus gas misalignment.]{Angular difference between the angular momentum vectors of cold ($T < 20000$K) gas within $0.1R_{200}$ and that of the stellar disc versus the tilting rate for low mass galaxies (black diamonds) and high mass galaxies (red squares), at $z=0.3$ (left),  $z=0.15$ (middle) and at $z=0$ (right). We find a strong correlation for the left and middle panels, and a slightly weaker correlation for the right panel. In the middle panel the dark and light grey shaded regions show the $1\sigma$ and $2\sigma$ confidence intervals, respectively.}
    \label{fig:vcgas_disc_dthetadt}
  \end{center}
\end{figure*}

\cite{roskar2010} showed that the angular momentum of the warp gives a good indication of the angular momentum of the corona.
As such, the misalignment of the warp and the stellar disc should shed light on the angular momentum that is being added to the disc from the corona, or directly through cold flows.
As our galaxies vary in disc scale-length and size we measure the angular momentum of all the gas out to $0.1 R_{200}$ to determine the angular momentum of the warp.
Figure \ref{fig:vcgas_disc_dthetadt} compares the angle between the angular momentum vectors of the cold gas ($T < 20000$K) and the stellar disc versus the tilting rate of the stellar disc, at $z=0.3$ (left), $z=0.15$ (middle) and $z=0$ (right).
We find a strong correlation between the cold gas misalignment with the tilting rate for $z=0.3$ and $z=0.15$, and a slightly weaker correlation at $z=0$.
A comparable analysis for cool gas ($T < 50000$K) gives similar results.

%%%%%%%%%%%%%%%%%%%%%%%%%%%%%%%%%%%%%%%%%%%%%%%%%%%%%%%%%%%%%%%%%%%%%%%%%%%%%%%%%%%

\subsection{Is the star formation rate a proxy for the tilting rate?}
\begin{figure}
  \begin{center}
    \includegraphics[width=\columnwidth]{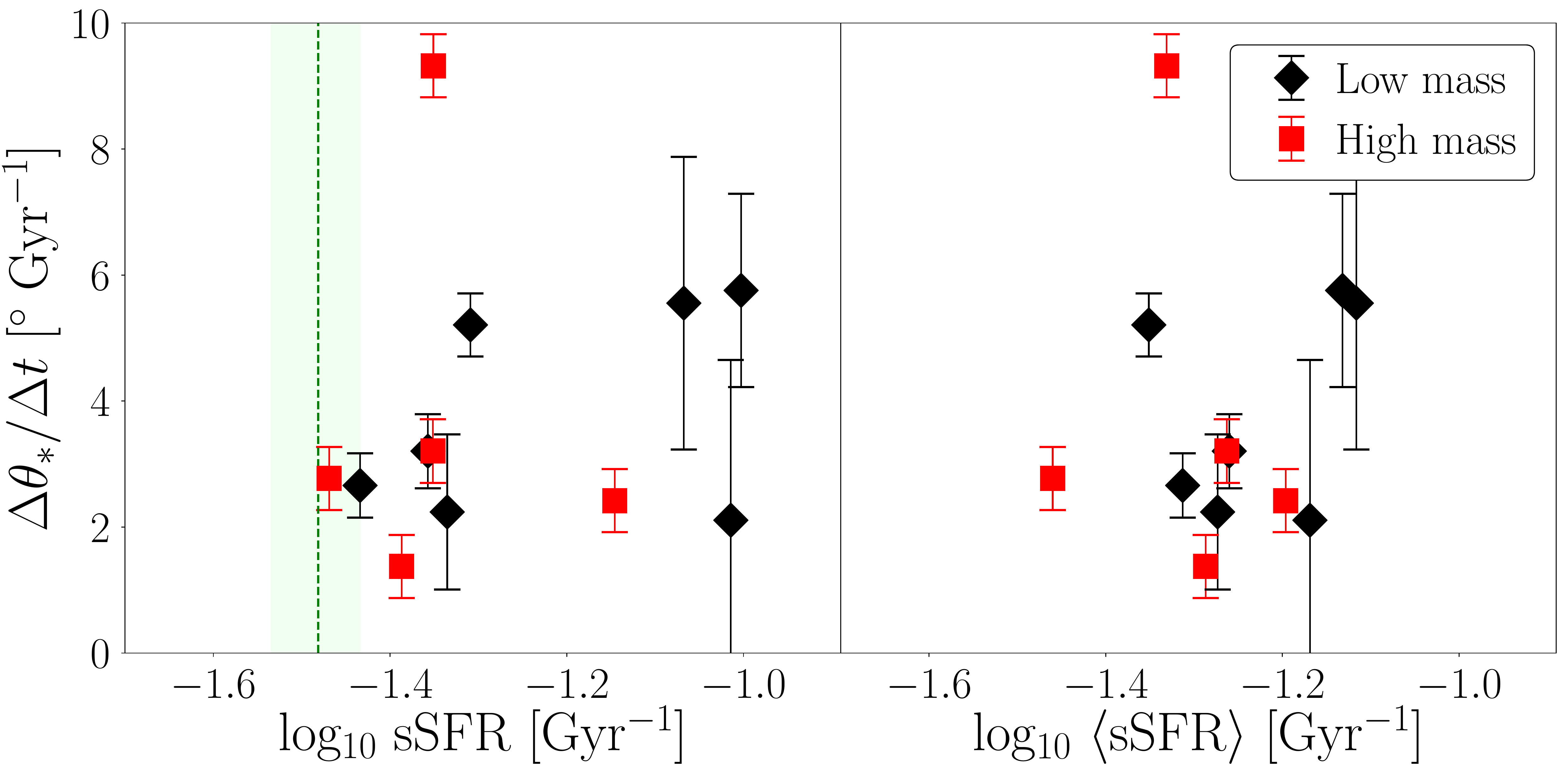}
    \caption[Tilting rate versus specific star formation rate.]{Left: specific star formation rate, at $z=0$, versus tilting rate of the stellar disc for low mass galaxies (black diamonds) and high mass galaxies (red squares). The (green) dashed line and shaded region show the specific star formation rate and the uncertainty of the MW calculated with the values of SFR and stellar mass from \cite{blandhawthorn2016}. Right: average specific star formation rate between $z=0.3$ and $z=0$. We find no correlation for either panel.}
    \label{fig:ssfrdthetadt}
  \end{center}
\end{figure}

Observationally detecting gas accretion or the alignment of the hot gas corona is extremely difficult.
However, as the galaxy requires this ongoing accretion of gas to fuel its star formation, the star formation rate itself is a proxy for the amount of gas reaching the disc.
We, therefore, measure the star formation rate (SFR) using the mass of all star particles born between subsequent time steps.
Since our subsample spans a wide range of masses, we use the specific star formation rate (sSFR), by dividing the SFR by the stellar mass within $R < 0.1R_{200}$, at the subsequent time step.
Figure \ref{fig:ssfrdthetadt} compares the sSFR at $z=0$ and averaged over the time range to the tilting rate of the stellar disc.
The green dashed line shows a recent estimate for the MW's current and averaged sSFR.
The left panel takes the value of $1.65 \pm 0.19 \Msun$ yr$^{-1}$ for the present day MW SFR, with the current stellar mass assumed to be $M_{*} \sim 5 \times 10^{10} \Msun$ \citep{blandhawthorn2016}.
We find no correlation between either the present day sSFR or the average sSFR and the tilting rate.
We do find a weak correlation between the tilting rate and the sSFR at $z=0.3$ as well as with the peak sSFR.

%%%%%%%%%%%%%%%%%%%%%%%%%%%%%%%%%%%%%%%%%%%%%%%%%%%%%%%%%%%%%%%%%%%%%%%%%%%%%%%%%%%

\subsection{Baryonic mass fractions}
\begin{figure}
  \begin{center}
    \includegraphics[width=\columnwidth]{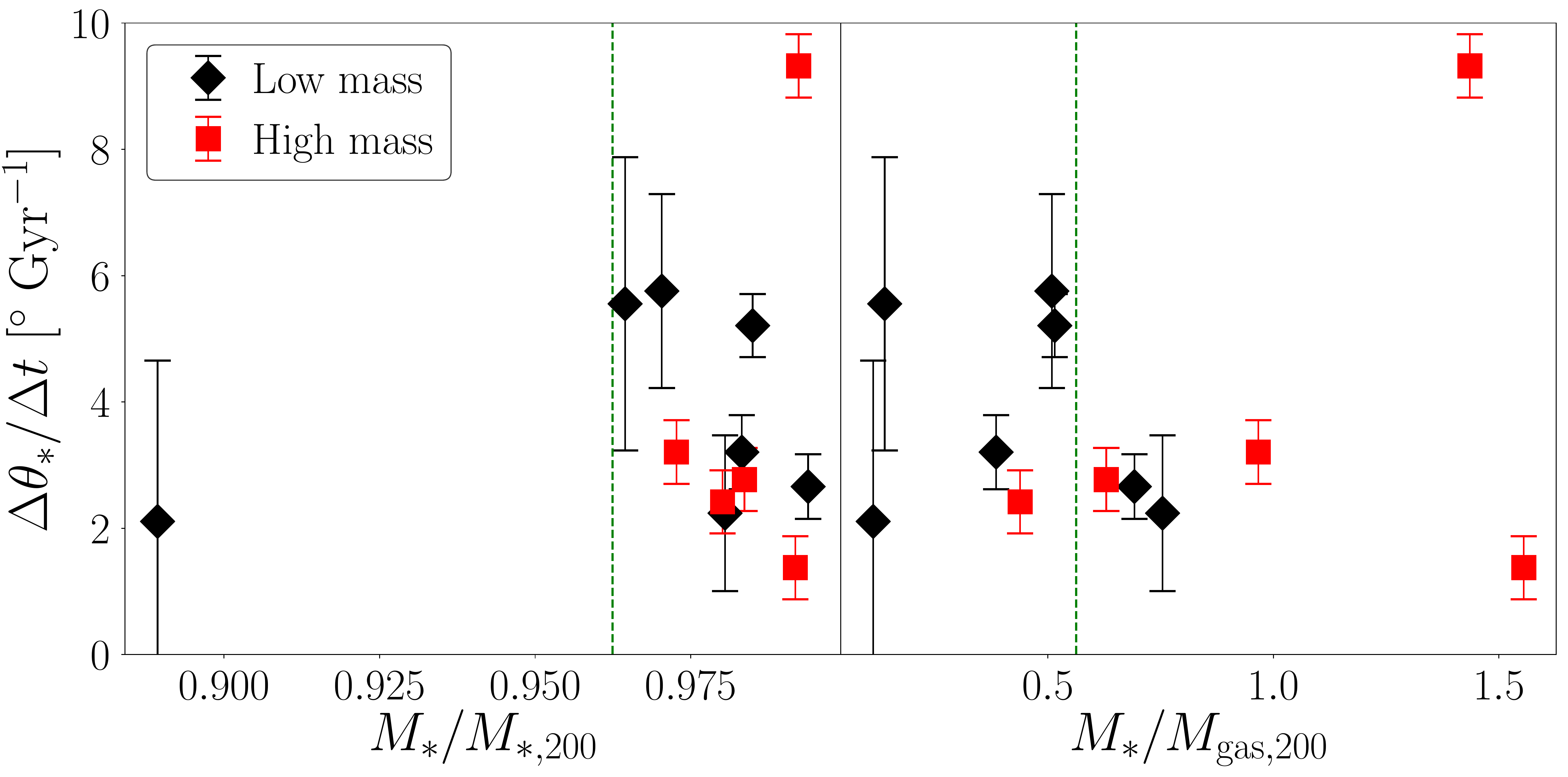}
    \caption[Tilting rate versus stellar mass ratios.]{Left: ratio of galaxy stellar mass out to 10 per cent of $R_{200}$ to total stellar mass inside $R_{200}$ versus tilting rate of the disc. Right: ratio of stellar mass out to $0.1R_{200}$ to total gas mass inside $R_{200}$ versus the tilting rate. The (green) dashed line on the right panel shows the stellar to gas mass ratio of the MW from \citet{blandhawthorn2016}. We find no correlation for either panel.}
    \label{fig:massfrac1}
  \end{center}
\end{figure}

The left panel of Figure \ref{fig:massfrac1} compares the ratio of stellar mass out to $0.1 R_{200}$ and total stellar mass out to $R_{200}$ to the tilting rate of the stellar disc.
The green line in this figure denotes an upper limit for the MW, assuming the LMC is the dominant stellar mass outside $0.1 R_{200}$ and has a virial mass of $M_{\rm LMC} \sim 2 \times 10^{11} \Msun$ \citep{gomez2015, kallivayalil2013, penarrubia2015}.
Determining the stellar mass from $M_{200}$ was done by assuming the LMC follows an observed abundance matching relation \citep{kravtsov2018}.
A value of $M_* \sim 5 \times 10^{10} \Msun$ was assumed for the stellar mass of the MW \citep{blandhawthorn2016}.
We find no correlation between the stellar mass ratio and the tilting rate of the stellar disc, indicating that the presence of nearby satellites within $R_{200}$ is not the main driver of the tilting rate; therefore, in the MW it is not likely that the LMC would be responsible for the tilting of the disc.
The right panel of Figure \ref{fig:massfrac1} compares the ratio of stellar (out to $0.1 R_{200}$) to total gas mass (out to $R_{200}$) against the tilting rate of the stellar disc. 
We find no correlation between either the mass of the hot gas corona or the sSFR and the tilting rate.
Moreover, we find no correlations between the tilting and either the stellar mass or the fractional change in baryonic/stellar mass.

%%%%%%%%%%%%%%%%%%%%%%%%%%%%%%%%%%%%%%%%%%%%%%%%%%%%%%%%%%%%%%%%%%%%%%%%%%%%%%%%%%%

\subsection{Angular Momentum}
\begin{figure}
  \begin{center}
    \includegraphics[width=\columnwidth]{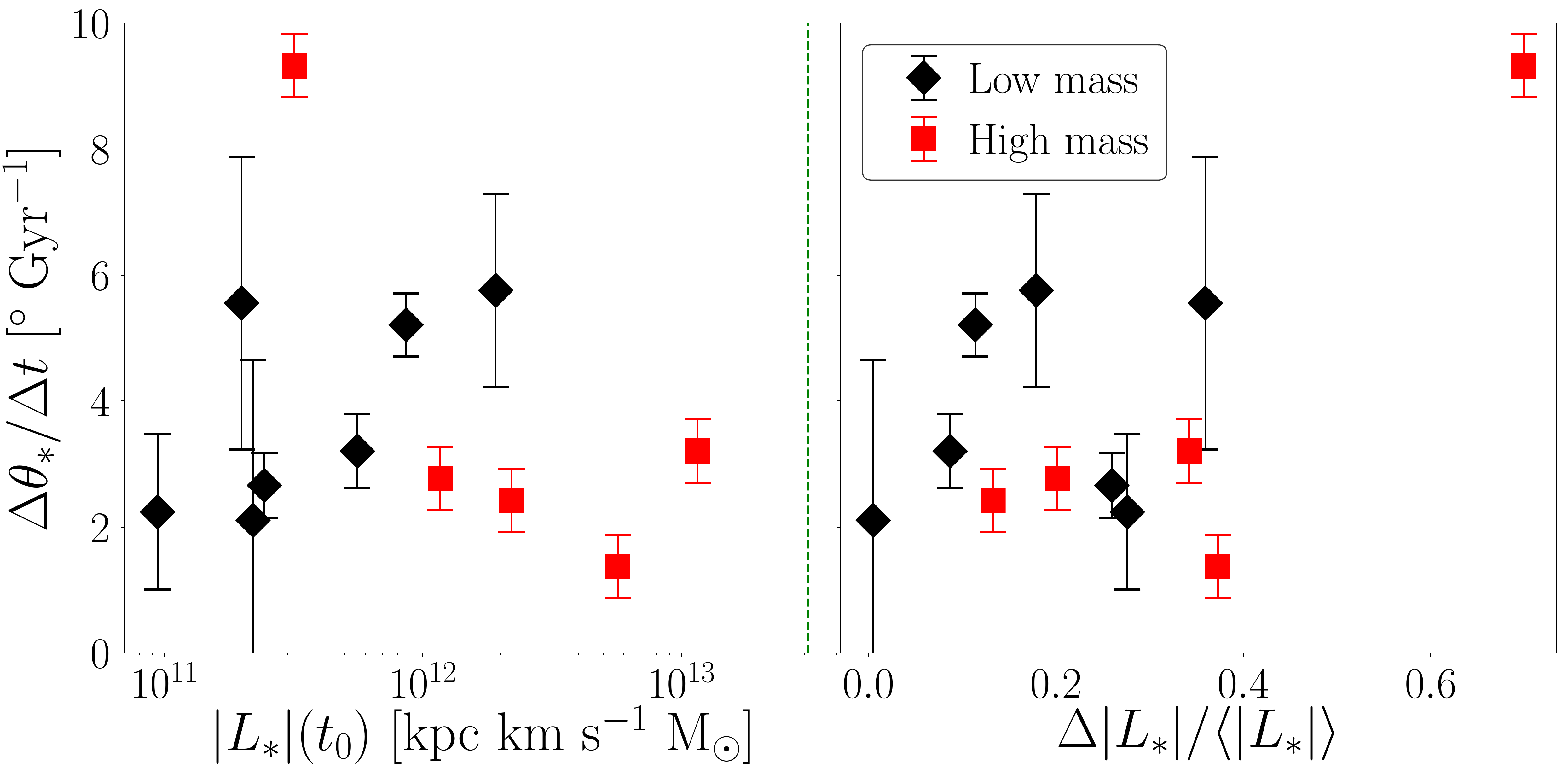}\\
    \includegraphics[width=\columnwidth]{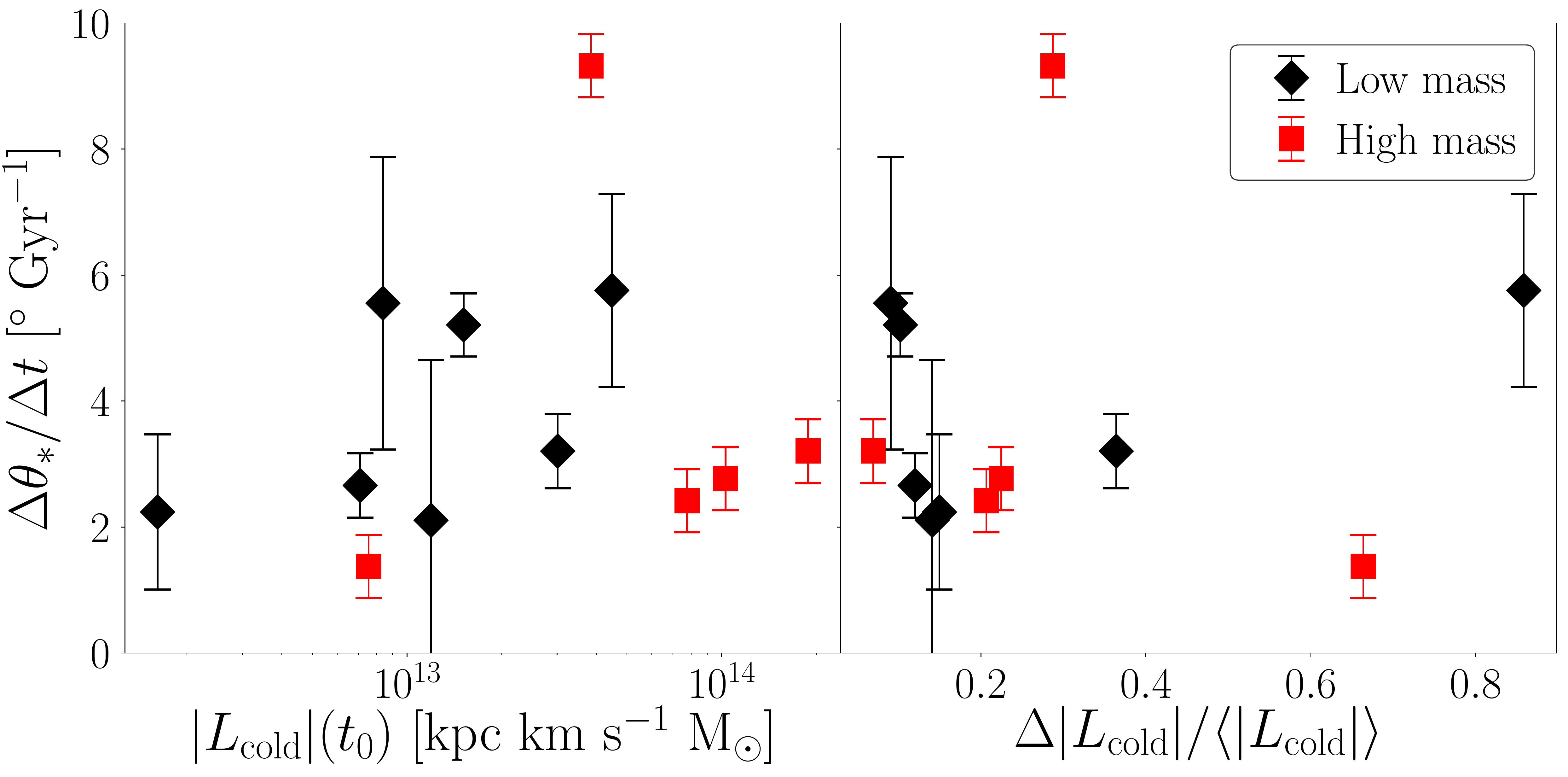}
    \caption[Tilting rate versus stellar angular momentum of the stellar and cool gas components.]{Left: the angular momentum of the stellar disc (top), and cold gas disc (bottom) at $z=0.3$. Right: the change in angular momentum between $z=0.3$ and $0$ divided by the mean angular momentum of the stellar disc (top), and cold gas disc (bottom). The (green) dashed line shows an estimate for the angular momentum of the MW's stellar disc. We find a weak anti-correlation for the top left panel, and a weak correlation for the top right panel. We find no correlation for the two bottom panels.}
    \label{fig:ldiscdthetadt}
  \end{center}
\end{figure}

Figure \ref{fig:ldiscdthetadt} shows the tilting rate of the stellar disc versus the angular momentum of the stellar disc (top) at $z=0.3$ (left), and its difference between $z=0.3$ and $z=0$ normalized by the mean angular momentum (right). 
A rough estimate for the angular momentum of the Milky Way's stellar disc is $|L| = M_d R_d V_c = 3.1 \times 10^{13} \Msun$ km s$^{-1}$ kpc (green dashed line), obtained by assuming a disc stellar mass to be $5 \times 10^{10} \Msun$ \citep{blandhawthorn2016}, all at the scale radius $R = 2.6$ kpc \citep{blandhawthorn2016}, and with a circular velocity of $240$ km s$^{-1}$ \cite{schoenrich2012, blandhawthorn2016}.
The bottom panels show the angular momentum of the cold gas disc ($T < 20000$K) at $z=0.3$ (left), and its change normalized by the average angular momentum over the same time period (right), versus the tilting rate of the stellar disc.
We find a weak anti-correlation between the amount of angular momentum in the stellar disc and the tilting rate, meaning galaxies with higher angular momentum tilt more slowly.
Moreover, we find a weak correlation between the normalized change in angular momentum of the stellar disc and its tilting rate.
This hints at a connection between the amount of angular momentum being gained by the disc and its tilting rate.
Therefore, the amount of angular momentum able to reach the disc might be an indicator of how fast the disc can tilt.
In the remaining two panels, we find no correlation.

%%%%%%%%%%%%%%%%%%%%%%%%%%%%%%%%%%%%%%%%%%%%%%%%%%%%%%%%%%%%%%%%%%%%%%%%%%%%%%%%%%%

\subsection{Multiple regression: double variable models}
To measure the statistical significance of including multiple variables we use weighted least squares (WLS) linear regression to determine the adjusted coefficient of determination ($\bar{R}^2$).
The coefficient of determination is given by,
$$R^2 = 1 - \frac{\sum_i (y_i - \hat{y}_i)^2}{\sum_i (y_i - \langle y \rangle)^2},$$
where $y_i$ is the observed value for each galaxy and $\hat{y}_i$ is the predicted value for each galaxy.
A value of $R^2 = 1$ would mean the model can explain all variability in the data, whereas, $R^2 = 0$ would mean the model fails to explain any of the variability.
When adding more parameters the value of $R^2$ can only remain the same or increase, therefore, we turn to the adjusted coefficient of determination, given as
$$\bar{R}^2 = 1 - (1 - R^2)\frac{n-1}{n-p-1},$$
where $p$ is the number of predictors, and $n$ is the number of observations.  
If we include an additional variable that does not improve the model, $R^2$ will stay the same, but $\bar{R}^2$ will decrease.
The addition of a statistically significant variable will increase both $R^2$ and $\bar{R}^2$.
As $p$ becomes bigger, the change in $R^2$ needed for the variable to be significant will also increase.
The WLS linear model is constructed with the {\sc statsmodels} package in Python, using $1/\epsilon_R^2$ for our weights.
To find the maximum value for $\bar{R}^2$ we use the forward selection form of stepwise regression.
For this method, we measure $\bar{R}^2$ for a model with just a constant term, then measure $\bar{R}^2$ after the addition of each variable. 
The variable that statistically improves the model the most is chosen as our first parameter.
We then measure $\bar{R}^2$ for a model with our first parameter and the inclusion of a second parameter from the remaining variables.
This process is repeated until the model no longer statistically improves or until we reach a specified number of parameters.
We consider all the baryonic predictors listed in Table \ref{tab:p_values} finding a best fitting, single parameter, linear model for the tilting rate of the stellar disc given by 
\begin{equation}\frac{\Delta \theta_*}{\Delta t} = 3.8 \theta(*,{\rm cold\;  gas}) + 1.6, \label{single_reg_model}\end{equation}
where $\theta(*,{\rm cold\;  gas})$ is the angle between the stellar disc and cold gas warp measured at $z=0.15$, with $R^2 = 0.74$ and $\bar{R}^2 = 0.71$.
This result is unsurprising, as the correlation between $\theta(*,{\rm cold\;  gas})$ and the tilting rate exhibits the largest p-value.
We compare this best fitting model to the best fitting model with two parameters, 
\begin{equation}\frac{\Delta \theta_*}{\Delta t} = 5.3 \theta(*,{\rm cold\;  gas}) - 3 M_{*}/M_{{\rm gas,200}} + 3.2, \label{mutli_reg_model}\end{equation}
where $M_{*}/M_{{\rm gas,200}}$ is the ratio of disc stellar mass to virial gas mass.
Equation \ref{mutli_reg_model} has a larger $R^2$ of $0.897$ and is a statistically significant improvement over Equation \ref{single_reg_model} with $\bar{R}^2 = 0.874$.
Therefore, a better fitting linear model uses two parameters $\theta(*,{\rm cold\;  gas})$ and $M_{*}/M_{{\rm gas,200}}$.

\begin{figure}
  \begin{center}
    \includegraphics[width=\columnwidth]{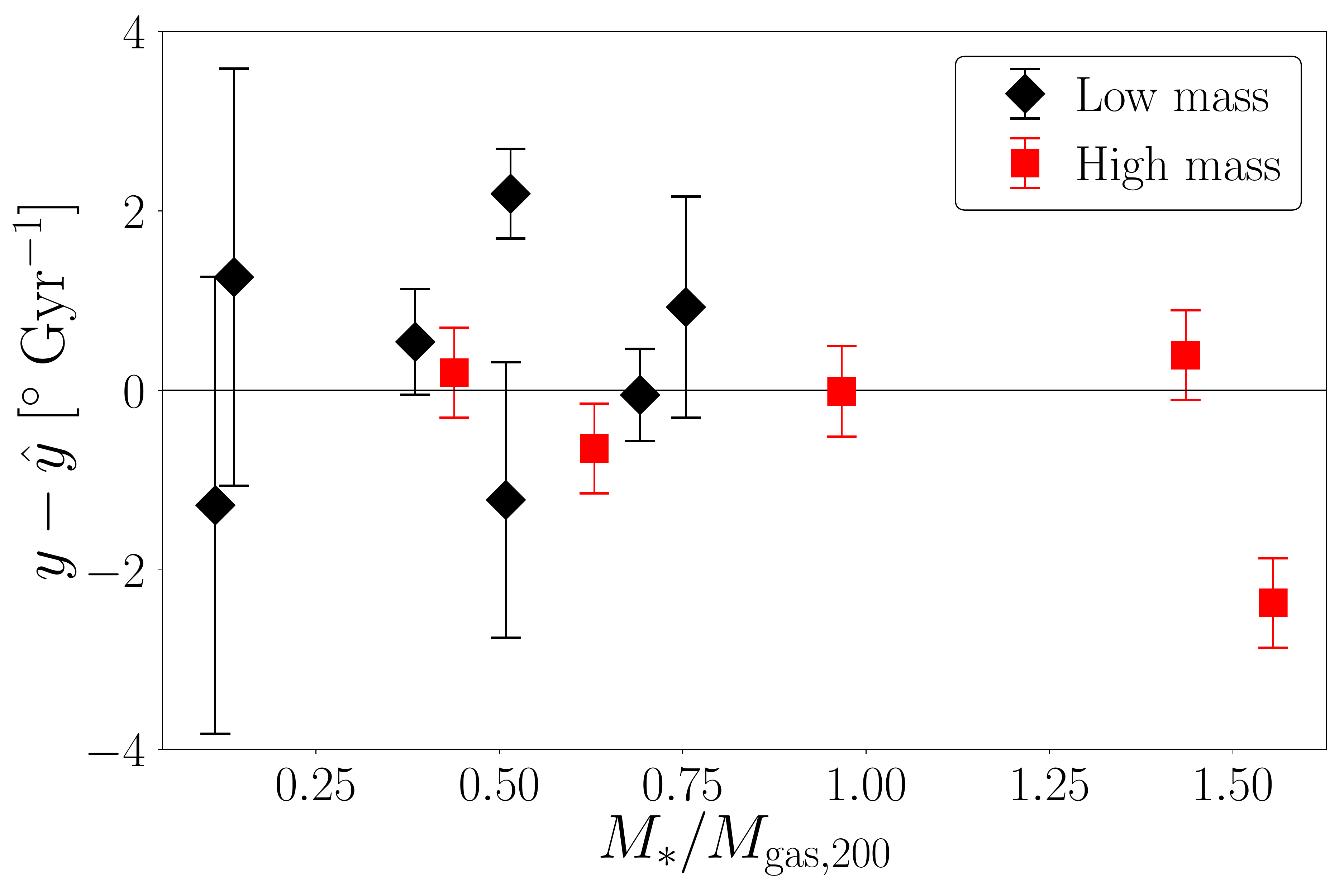}
    \caption[Best fitting model residuals.]{Residuals between the expected values for the tilting rate ($\hat{y}$) from the best fitting linear model and the measured tilting rates ($y$) against the ratio of disc stellar mass to virial gas mass. The (black) dashed line shows $y - \hat{y} = 0$. The linear model is given by $\hat{y} = 3.7 \theta(*,{\rm cold\;  gas}) + 1.6$.}
    \label{fig:model_residuals}
  \end{center}
\end{figure}

Figure \ref{fig:model_residuals} shows the residuals between the expected values for the tilting rates using Equation \ref{single_reg_model} and the measured tilting rates against the ratio of disc stellar mass to virial gas mass.
We see a negative trend which, due to the increase in $\bar{R}^2$ between Equations \ref{single_reg_model} and \ref{mutli_reg_model}, is statistically significant.

We used the same method on the dark matter predictors listed in Table \ref{tab:p_values}. We find no model that can predict the tilting rate better than $R^2 = 0.47$ and $\bar{R}^2 = 0.35$.
Therefore, no model using dark matter predictors is able to match the tilt rate as well as models using gas predictors.

%%%%%%%%%%%%%%%%%%%%%%%%%%%%%%%%%%%%%%%%%%%%%%%%%%%%%%%%%%%%%%%%%%%%%%%%%%%%%%%%%%%
% Discussion & Conclusions
%%%%%%%%%%%%%%%%%%%%%%%%%%%%%%%%%%%%%%%%%%%%%%%%%%%%%%%%%%%%%%%%%%%%%%%%%%%%%%%%%%%
\section{Discussion and Conclusions}
\label{sec:conclusions}
We have presented the tilting rates for galaxies in a suite of hydrodynamical cosmological simulations and tested for correlations with various properties.
As in E17, we find that all our galaxies tilt with rates that would be detectable by {\it Gaia}, assuming a detection limit of $0.27 \degrees$ Gyr$^{-1}$ \citep{perryman2014}.
We find error-weighted average tilting rates for our subsample of $3.8 \pm 2.3 \degrees$ Gyr$^{-1}$.
If we restrict our subsample to high mass galaxies, we find error-weighted average tilting rates of $3.8 \pm 2.8 \degrees$ Gyr$^{-1}$.

%%%%%%%%%%%%%%%%%%%%%%%%%%%%%%%%%%%%%%%%%%%%%%%%%%%%%%%%%%%%%%%%%%%%%%%%%%%%%%%%%%%

\subsection{The lack of a dark matter influence}
Using cosmological pure {\it N}-body simulations \cite{yurinspringel2015} reported tilting rates for their stellar discs greater than the tilting rates presented here.
In such a scenario the tilting of the stellar disc is driven by the torques, dynamical friction and interactions imposed by the tilting host DM halo and surrounding substructure.
To test whether torquing by the dark halo drives the tilting of the stellar disc, we compare the tilting rate of the stellar disc to the axis ratios, $b/a$ and $c/a$ as well as the triaxiality of the halo and find no correlations.
We also compare the tilting rate of the dark halo's angular momentum to that of the stellar disc, finding strong correlations for the tilting rates within $0.1 R_{200}$.
However, at such low radii, the halo's angular momentum is dominated by the stellar disc. 
Therefore, it is more plausible that the tilting of the dark matter is driven by the tilting of the stellar disc, not vice versa.
Moreover, we compared the tilting rate of the disc to the misalignment with respect to each axis of dark halo, finding no correlations.

We investigate the correlation found in E17 that galaxies in denser regions tilted with higher rates. 
We find no such correlation between the local overdensity and the tilting rate at any radius between 4 and 8 Mpc.
This difference is most likely due to selection criteria imposed on the NIHAO sample, which were absent in the galaxies presented in E17.
E17 did not choose galaxies to be preferentially in the field, resulting in some galaxies residing in cluster/group environments with overdensities as large as $\rho / \rho_{\rm crit} \sim 10$, whereas the largest values for NIHAO are $\sim 2$.
Dense environments could greatly affect the tilting of galaxies due to the higher frequency of tidal interactions.

Finally, we tried to combine the dark matter predictors into a single model using multiple regression, but were unable to create a model that can predict the tilting rate with an $\bar{R}^2 > 0.35$, far worse than models using gas predictors.
Combining these five results, it becomes hard to argue that the torques imposed by the dark matter are the primary drivers for disc tilting in the presence of gas.

Although the majority of galaxies in NIHAO are star forming, one of the galaxies in our sample has sSFR lower than the MW.
This galaxy also has the lowest tilting rate, $0.68 \pm 0.23 \degrees$ Gyr$^{-1}$, and its stellar disc is aligned to within $8 \degrees$ of the minor axis of its dark halo at $z=0$.
This makes its disc the most aligned with the minor axis of any galaxy in our sample, with the lowest sSFR rate of $0.022$ Gyr$^{-1}$.
Therefore, in the absence of gas, the dark halo can begin to drive disc tilting, as D15 predicted.

%%%%%%%%%%%%%%%%%%%%%%%%%%%%%%%%%%%%%%%%%%%%%%%%%%%%%%%%%%%%%%%%%%%%%%%%%%%%%%%%%%%

\subsection{The strong influence of gas}
The angle between the angular momentum vectors of the gas warp and the stellar disc, at both $z=0.3$ and $z=0.15$, correlates very strongly with the tilting rate of the disc.  The angular difference at $z=0$ provides a weaker correlation with the tilting rate.
Since we calculate the tilting rate between $z=0.3$ and $z=0$, the misalignment of the warp at $z=0$ should not be a good indicator of the tilting rate since $z=0.3$.
Of all the correlations explored between the gas and the stellar disc, these are the strongest, suggesting that the warp may be a good indicator of the tilting rate.
We improved on this correlation by building a linear multiple regression model.
We found that the combination of angular misalignment of the cold gas warp at $z=0.15$ and the stellar disc to total gas ratio were the best predictors finding $R^2 = 0.897$ and $\bar{R}^2 = 0.87$.

We find a weak anti-correlation between the amount of angular momentum in the stellar disc and its tilting rate.
This is not unexpected, as discs with greater angular momentum will be harder to tilt.
When we take the change in the magnitude of the disc's angular momentum and normalize by its average over the same time interval, we also find a weak correlation.
Therefore, galaxies that have gained fractionally more angular momentum exhibit faster tilting.

We compared the tilting rate of the disc to the sSFR measured at $z=0$, and at $z=0.3$, finding only a weak correlation for sSFR at $z=0.3$.
We also compared the average sSFR since $z=0.3$ and the peak star formation rate, still finding no correlation.

To determine the effect of the hot ($T > 50000$K) gas corona, we test for correlations between the misalignment and tilting rate of the angular momentum of the hot gas corona and the stellar disc's tilting rate. 
We find no correlations between the tilting rates of the two different components and the angular misalignment.
 We also tested this result using a higher temperature for the cutoff ($T > 100000$K), finding similar results.

D15 reported that redder galaxies tended to be aligned with the minor axis of their dark halo.
We measured the alignment between the stellar disc and the principal axes of the dark halo for the predominantly blue galaxies within the NIHAO sample.
We find no preferential alignments, in agreement with D15.

We have presented the tilting rates of galaxies in cosmological simulations and a possible link to the warp and the role of misaligned gas accretion.
As a next step, we will demonstrate, directly, the role of misaligned gas on the tilting of stellar discs.

%%%%%%%%%%%%%%%%%%%%%%%%%%%%%%%%%%%%%%%%%%%%%%%%%%%%%%%%%%%%%%%%%%%%%%%%%%%%%%%%%%%
% Acknowledgements
%%%%%%%%%%%%%%%%%%%%%%%%%%%%%%%%%%%%%%%%%%%%%%%%%%%%%%%%%%%%%%%%%%%%%%%%%%%%%%%%%%%
\section*{Acknowledgements}
SWFE would like to thank the Max-Planck-Institut f\"{u}r Astronomie, Heidelberg for their hospitality, as well as Aaron Dutton, Benjamin MacFarlane and Min Du for useful conversations.
VPD is supported by STFC Consolidated grant \#~ST/R000786/1.
TB acknowledges support from the  Sonderforschungsbereich SFB 881 `The Milky Way System' (subproject A2) of the German Research Foundation (DFG).
Simulations were carried out on the High Performance Computing resources at New York University Abu Dhabi.
This work made use of the {\sc pynbody} package \citep{pynbody} to analyse simulations and the {\sc python} package {\sc matplotlib} \citep{matplotlib} to generate all the figures for this work. 
The data analysis for this work was carried out using the {\sc python} packages {\sc scipy}, {\sc numpy}, {\sc ipython} and {\sc jupyter} \citep{scipy,numpy,ipython,jupyter}.

%%%%%%%%%%%%%%%%%%%%%%%%%%%%%%%%%%%%%%%%%%%%%%%%%%%%%%%%%%%%%%%%%%%%%%%%%%%%%%%%%%%
% References 
%%%%%%%%%%%%%%%%%%%%%%%%%%%%%%%%%%%%%%%%%%%%%%%%%%%%%%%%%%%%%%%%%%%%%%%%%%%%%%%%%%%
\bibliographystyle{mnras}
\bibliography{ms}

%%%%%%%%%%%%%%%%%%%%%%%%%%%%%%%%%%%%%%%%%%%%%%%%%%%%%%%%%%%%%%%%%%%%%%%%%%%%%%%%%%%

\bsp
\label{lastpage}
\end{document}